## Effects of a tungsten addition on the morphological evolution, spatial correlations, and temporal evolution of a model Ni-Al-Cr superalloy

Chantal K. Sudbrack[1,2], Tiffany D. Ziebell[1,3], Ronald D. Noebe[4] and David N. Seidman[1,5]*

[1]Department of Materials Science and Engineering,
Northwestern University, 2220 Campus Drive, Evanston, IL 60208 USA
[2] Materials Science Division, Argonne National Laboratory, Argonne, IL 60439 USA
[3]Department of Materials Science and Engineering, Massachusetts Institute of Technology, 77 Massachusetts Avenue, Cambridge, MA 02139 USA
[4]NASA Glenn Research Center, 21000 Brookpark Rd., Cleveland, OH 44135, USA
[5]Northwestern University Center for Atom-Probe Tomography, 2220 Campus Drive, Evanston, IL 602808 USA
*Corresponding author: d-seidman@northwestern.edu, 847-491-4391

### Abstract

The effect of adding 2 at.% W to a model Ni-Al-Cr superalloy on the morphological evolution, spatial correlations and temporal evolution of γ'(L1$_2$)-precipitates at 1073 K is studied with scanning electron microscopy and atomic force microscopy. Adding W yields a larger microhardness, earlier onset of spheroidal-to-cuboidal precipitate morphological transition, larger volume fraction (from ~20 to 30%), reduction in coarsening kinetics by one third and a larger number density ($N_v$) of smaller mean radii ($<R>$) precipitates. The kinetics of $<R>$ and interfacial area per unit volume obey $t^{1/3}$ and $t^{-1/3}$ relationships, respectively, which is consistent with coarsening driven by interfacial energy reduction. The $N_v$ power law dependencies deviate, however, from model predictions indicating that a stationary-state is not achieved. Quantitative analyses with precipitate size distributions, pair correlation functions, and edge-to-edge interprecipitate distance distributions gives insight into 2D microstructural evolution, including the elastically driven transition from a uniform γ'-distribution to one-dimensional <001>-strings to eventually clustered packs of γ'-precipitates in the less densely packed Ni-Al-Cr alloy.

## I. Introduction

The widespread use of $\gamma'(L1_2)$-precipitate strengthened Ni-base superalloys as a turbine blade material in high temperature commercial and military jet engines and land-based gas turbines warrants extensive study of the microstructural evolution and $\gamma'$-coarsening behavior in model $\gamma(fcc)/\gamma'$ alloys, such as Ni-rich Ni-Al, Ni-Al-Cr and Ni-Al-Mo alloys [1-13]. In these alloys, elastic anisotropy leads to coherent precipitates evolving from spheres-to-cubes to arrays of cubes or to doublets or octets of smaller precipitates [1,8,14-17], as well as face-to-face cube alignment along the elastically soft <100>-crystallographic directions [1,2]. In all cases, the morphology is dictated by the minimization of the interfacial free energy and elastic free energy arising from lattice parameter misfit, while the spatial correlations are controlled by elastic and diffusional interactions among neighboring precipitates. To improve its high temperature performance, modern commercial Ni-base superalloys contain as many as 12 or 13 microalloying elements [18,19]. High-melting point refractory element additions, such as Nb, Re, Ru, W, and Mo, are particularly effective in improving creep resistance, alloy strength, and operating temperature [20]. The chemical and microstructural $\gamma/\gamma'$ decomposition that controls these properties is strongly dependent on the complex interplay between elastic interactions of misfitting $\gamma'$-precipitates and rate-controlling diffusion of these refractory elemental additions with small diffusivities. The research reported herein is part of a systematic study [21-24] to understand the roles of small additions of refractory elements on the $\gamma/\gamma'$ decomposition when a model, nominally Ni-10 Al-8.5 Cr (at. %), superalloy is aged at 1073 K. Herein, the influence of adding 2 at.% W is considered. Of all possible candidates for solid-solution strengthening, W has the highest melting temperature and is among the slowest diffusers in Ni [20]; however, relative to other refractory additions it has the largest molar volume compared to Ni, 36.93 % greater than Ni [25], which can significantly impact the $\gamma/\gamma'$ lattice parameter mismatch that drives the elastic interactions.

We reported previously [21] on the elemental partitioning behavior, evolving phase chemical composition, and γ/γ' heterophase interfacial chemistry at 1073 K as measured with conventional atom-probe tomography (APT) for the Ni-Al-Cr and Ni-Al-Cr-W alloys analyzed further here. In addition to details on the evolving chemical composition, the sample volumes analyzed in 3D with conventional APT (~ 15 x 15 x 100 $nm^3$) yield an understanding of the γ'-precipitates' morphology, dimensions, and growth rate. Post-mortem analyses with conventional APT reveals it is not possible to suppress γ'-precipitation that occurs during a water-quench after the alloys are homogenized, which is a result of the large solute supersaturations in these two alloys. The as-quenched alloys contain γ'-precipitates, 5-15 nm in three-dimensional radius, that do not exhibit a significant degree of coagulation and coalescence. This result is in contrast to a high degree of γ' coagulation and coalescence observed for an as-quenched quaternary alloy with 2 at.% Re [23,24,26]. In addition to reducing the γ'-precipitate growth and coarsening rate, the W additions partition preferentially to the γ'-precipitates with no evidence of nonmonotonic (confined) segregation of W at the γ/γ'-heterophase interface. Adding 2 at.% W also leads to a stronger partitioning of Al to γ'-precipitates and of Cr to the γ-matrix than measured for the Ni-Al-Cr alloy. At the shortest aging times (< 1 h), the γ'-precipitates exhibit concentration gradients in W in which their cores contain a smaller W concentration, approximately 2.0 at.% W, than the concentration within the regions immediately adjacent to the γ/γ' heterophase interface, approximately 2.5 at.% W. We note that the two alloys exhibit only small changes in chemical compositions after 4 h. This is when the alloys have reached a quasi-stationary state coarsening regime, i.e. when change in matrix supersaturation with time is approximately zero. Our present microstructural evaluation with scanning electron microscopy (SEM) and atomic force microscopy (AFM) focuses on the evolution of precipitate morphologies and precipitate spatial correlations in the quasi-stationary state coarsening regime (*t* > 4 h), where the elemental partitioning between the γ- and γ'-phases is nearly constant at precipitate dimensions that are reasonably well-resolved by these microscopic techniques. Thus permitting the coarsening behavior to be modeled accurately with predictions for multi-component alloys [27,28].

## II. Experimental Procedures

To produce master ingots, high purity constituent elements (99.97 Ni wt. %, 99.98 Al wt. %, 99.99 Cr wt. %, and 99.98 wt. % W) were induction melted under an Ar atmosphere to minimize oxidation, and then chill cast in a cylindrical copper mold, 19 mm in diameter by 86 mm long, with a conical hot top section to accommodate shrinkage within the casting. The cast ingots were homogenized at 1573 K for 20 hours, and the Ni-Al-Cr and Ni-Al-Cr-W alloys were further solutionized for 0.5 h at 1273 K and 1363 K, respectively, in the single-phase region (approximately 90 K above solvus [21]) to minimize γ'-precipitate formation during the subsequent water quenching. As-quenched sections were aged at 1073 K for times ranging from essentially zero (the as-quenched state) to 264 h, followed again by a rapid water-quench from the aging temperature.

Bulk chemical analysis of the two alloys was performed using an Applied Research Laboratories Model 3560 inductively coupled plasma (ICP) atomic-emission vacuum spectrometer. ICP atomic-emission spectroscopy was also used to determine the composition of γ- and γ'-phases directly from specimen sections aged for 264 h; the two phases were separated electrolytically by standard phase-extraction techniques. The results of these chemical analyses are presented in Table 1. Phase extraction was accomplished by anodic dissolution of the γ-matrix phase with a 1:1 aqueous solution of citric acid and ammonium nitrate at constant current density. By comparing the weight of the extracted γ'-precipitates with the overall section weight, the volume fraction of the γ'-precipitates, $\phi(t = 264$ h), is measured to be 0.189 and 0.308 for the Ni-Al-Cr and Ni-Al-Cr-W alloys, respectively. Utilizing THERMO-CALC [29] and the Ni-DATA v.4 thermodynamic database [30,31], the calculated equilibrium values of $\phi$ at 1073 K are 0.179 and 0.332 for the alloys, respectively. These calculated values are in good agreement with our experimental values, which suggests that the phase composition and $\phi$ measurements obtained from the 264 h aged specimens are reasonable measures of the equilibrium values, $\phi^{eq}$. Tracer diffusivities at 1073 K [32] reported in Table 1 were calculated for face-centered cubic (fcc) alloys employing the thermodynamic database developed by Saunders [31] and the mobility database developed by Campbell [33] for Ni-based superalloys.

For the metallographic investigation of the grain structure, samples were mounted perpendicular to the ingot length, mechanically polished to 1 μm, and cleaned ultrasonically in a solution of 20 parts deionized $H_2O$ with 1 part of a Liqui-Nox Critical-Cleaning liquid detergent and dried between successive polishing steps. Samples were then etched chemically with "waterless Kahling's solution" consisting of a mixture of 2 gm $CuCl_2$, 40 mL HCl, and 80 mL ethanol. Vickers microhardness for a 500 g load sustained for 5 s was measured on mounted samples polished to 1 μm. The average value of fifteen independent microhardness measurements made on several grains was determined and converted from $kg/mm^2$ to MPa by multiplying by the acceleration of gravity, 9.807 $m/s^2$.

It was necessary to obtain a higher quality polish on mounted samples for the samples investigated with SEM and AFM. To achieve nanoscale flatness, 1 μm-polished samples were polished with a 0.3 μm alumina abrasive (a dime-sized amount was placed in the center of the micro-cloth and diluted with either deionized $H_2O$ or with a solution of 20 parts deionized $H_2O$ to with 1 part Liqui-Nox Critical-Cleaning liquid detergent to lubricate the wheel) followed by a Mastermet 0.02 μm colloidal silica suspension diluted 1:5 times with deionized $H_2O$. A mixture of HCl, distilled water, and $K_2S_2O_5$ etches effectively the matrix phase in Ni-Al alloys [34], and is an effective etchant for both the ternary and quaternary alloys. For the Ni-Al-Cr alloy, carefully cleaned and dried samples were etched with a 100 mL HCl / 100 mL deionized $H_2O$ / 0.5 g $K_2S_2O_5$ solution for up to 30 seconds, depending on the aging state. For the alloy containing W, 100 mL HCl / 100 mL deionized $H_2O$ / 1 g $K_2S_2O_5$ mixture was used for similar durations, and the overall etching rate for both mixtures was roughly 1 nm $s^{-1}$.

The γ'-nanostructure was imaged with a LEO Gemini 1525 field-emission gun SEM operated at 3 kV with a 20 - 30 μm aperture and 8 to 11 mm working distances. SEM images were taken from several grains for each aging time, and each micrograph contained 300 to 850 γ'-precipitates. The γ'-nanostructure was also imaged with contact mode AFM [35,36] using a SiN AFM tip with an apex radius of 25 nm. The polish relief depth was measured during AFM imaging utilizing three to five independent line trace profiles from several regions of the sample.

Measurements of the cross-sectional microstructure of polish-relief surfaces are determined directly from SEM images and are based on 1000 to 3500 γ'-precipitates for each aging time. Individual precipitates within the images were traced by hand using commercial software (ADOBE PHOTOSHOP 6.1), and the dimensions of the traced areas were determined utilizing NIH IMAGE 1.62, a freeware image processing program, which fits the precipitate cross-sectional areas to a rectangular box or to a best-fit ellipse. The caliper distance, $H'$, is determined from the average of the rectangular box's two dimensions, where the prime notation is a convention adopted from image analysis of finite sections with one transparent phase and one opaque phase [37]. The two-dimensional radius, $R_{PS}$, is calculated from the traced areas using a circular-equivalence, that is $R_{PS} = \sqrt{area/\pi}$, where the measurements are approximately equivalent to those determined from a planar section (PS). Individual outlines of the precipitates' edges were labeled with a unique grayscale value employing a subroutine, BinaryLabel8, contained in a morphology collections plug-in [38] developed for IMAGEJ, the successor to NIH IMAGE, using version 1.37. From the labeled outlines, a two-dimensional edge-to-edge interprecipitate distance, $\lambda_{e-e}^{PS}$, is determined using Pycluster [39], a clustering module for Python, with a Python program adapted from one written for 3-dimensional APT reconstructions [40]. The aspect ratio for each cross-sectional area is determined from the ratio of the major axis to the minor axis of its best-fit-ellipse, where a lower threshold value of 1.75 is used to delineate the percentage of precipitates that are coalesced, $f$. The points per unit length, $P_l'$, were obtained from the intersections of a cycloid grid with the γ/γ' interfaces. The error in tracing is given by twice the physical length of an image pixel, which ranges from 1 - 5 nm depending on the image magnification, and gives a reasonable estimate for the error in <H'> and $\lambda_{e-e}^{PS}$. The error in the areal density of precipitates, $N_a{}'$, as well as in $P_l{}'$, is determined directly from counting errors, while the error in areal fraction, $A_a{}'$, and in $R_{PS}$ are based on standard error propagation methods [41] of the counting and tracing errors.

For the two-dimensional precipitate size distributions (PSD$_{PS}$) and interprecipitate distance distributions (IDD$_{PS}$), the number of γ'-precipitates is plotted in a histogram versus a scaled $R_{PS}/<R_{PS}>$ and $\lambda_{e-e}^{PS}/<R_{PS}>$, respectively, where the interval size varies between 0.1 and 0.2 of the scaled quantity. The distributions are normalized by dividing the count in an individual interval with the total count; that is, the total number of precipitates analyzed, and with the scaled dimension of the interval width. Similarly, the two-dimensional pair-correlation functions (PCF$_{PS}$) are determined on lightly etched planar sections, and are referred to as PCF$_{PS}$. The PCF$_{PS}$ is defined as the ratio of the number of precipitates per shell area, $N_a{}'(r_j)$, whose centers are within a circular shell at a radial distance, $r_j$, of a defined thickness, to the total number of precipitates per area, $N_a{}'$, as given by:

$$PCF(r_j)_{PS} = \frac{N_a{}'(r_j)}{N_a{}'}. \tag{1}$$

As $r_j \rightarrow \infty$, $PCF_{PS} \rightarrow 1$ for nearly all spatial distributions.

## III. Results

### 3.1. Metallography and hardness measurements

Figure 1 contains representative images of the grains from aged samples of Ni-9.8 Al-8.3 Cr (at.%) and Ni-9.7 Al-8.5 Cr-2.0 W (at.%) alloys. For the Ni-Al-Cr alloy, the grains are coarse, nearly equiaxed, and slightly twinned. Smaller grains with a 0.5 mm diameter ornament the ingot's edges, while the central region of the ingot contains larger grains whose diameter is typically 2 - 3 mm. The variations in grain contrast when imaged with polarized light suggest that the grains are oriented in different crystallographic directions. In the Ni-Al-Cr-W alloy, the grain structure is less coarse. Grains are elongated radially with a radial extension that is 2 to 6 times greater than their width (~ 0.5 mm). In this case, there is no appreciable contrast between grains when imaged with polarized light. Field-ion microscopy experiments on the aged alloys [21] confirm that the grains in the Ni-Al-Cr alloy are oriented in different directions along the ingot length, while the grains in the Ni-Al-Cr-W alloy have a strong <110>-texture; that is, perpendicular to the plane of view shown in Figure 1b.

Microhardness measures indirectly the dynamics of the precipitation sequence through hardness changes. It is a function of the spatial distribution of precipitates, their dimensions, as well as their $N_v$ value, where peak microhardness corresponds to the optimal balance between the mean γ'-precipitate $<R(t)>$ and $N_v(t)$ with aging. For the Ni-Al-Cr alloy, the microhardness for $t = 0 - 264$ h varies over a narrow range, 2.15 GPa to 2.5 GPa at 1073 K (Figure 2). Peak microhardness is achieved at 4 h, after which the microhardness decreases steadily. For the Ni-Al-Cr-W alloy, the 1073 K values are 40 to 50% greater than the Ni-Al-Cr measurements, which likely results from the denser packing of the γ'-precipitates and added solid-solution strengthening with the W addition. Similar to the Ni-Al-Cr alloy, the peak hardness (3.5 GPa) is achieved after 4 h; however, the Ni-Al-Cr-W alloy retains this peak up to 64 h of aging. For the final aging time of 264 h, the microhardness decreases to $3.27 \pm 0.10$ GPa after the plateau.

### 3.2. Nanostructural and morphological development in 2D

Figure 3 displays SEM images of the γ'-precipitates (light) protruding above a chemically etched γ-matrix (dark) at a constant magnification and provides a cross-sectional view of the three-dimensional nanostructure as it evolves temporally at 1073 K within the Ni-Al-Cr alloy (*upper four-square panel*, (a-d)), and within the quaternary alloy with a 2 at. % W addition (*lower four-square panel*, (e-h)). The four aging times studied increase by multiples of four, to double the root-mean-squared (RMS) diffusion distance for each time step. Figure 4 displays a typical topological AFM image of a γ'-nanostructure (light contrast) used to ascertain the depth of polish-relief. The etching procedure used has the requisite control to avoid precipitate fall-out and over etching. Polish-relief depths ($h$) range from several to tens of nanometers and are 3 - 12 times greater than $<H'>$, with exception of the 4 h aging time for the Ni-Al-Cr alloy, which was slightly over etched (Table 2).

For the Ni-Al-Cr alloy (Figure 3 (a-d)), the γ'-precipitates for specimens aged for 4 and 16 h appear spheroidal, isolated, and uniformly distributed throughout the γ-matrix. After 64 h, the precipitates become slightly cuboidal; that is, the corners of cubes are significantly rounded, marking the onset of the elastically-induced sphere-to-cube morphological transition. For the final aging time of 264 h, the corners of the cuboids become more sharply defined as their faces flatten, establishing a more-faceted cube-like morphology. For the Ni-Al-Cr-W alloy (Figure 3 e - h), the morphological evolution is influenced by the denser packed γ'-distribution. As shown in Table 2, the $<\lambda_{e-e}^{PS}>$ values are 1.3 to 2.6 times smaller than the corresponding Ni-Al-Cr values (Table 2). For specimens aged for 4 and 16 h, the $<\lambda_{e-e}^{PS}>$ is only 9.1 ±2.0 nm and 13.2 ± 2.5 nm, respectively, which allows coagulation and coalescence among the closely-spaced precipitates that results in some irregularity in the spheroidal morphology (e.g., L-shaped). By 64 h of aging, $<\lambda_{e-e}^{PS}>$ increases to ~ 24 nm. The γ'-precipitates are clearly faceted cuboids and the spheroidal irregularities are no longer observed. For $t \geq 64$ h, approximately 5 % of the γ'-precipitates (Table 2) have an aspect ratio > 1.75 and exhibit a nonequiaxed rectangular parallelepiped morphology, that is common to later stages of coarsening in Ni-Al alloys [1,2].

During the earliest stages of decomposition, the nucleation current (nuclei m$^{-3}$ s$^{-1}$) is extremely sensitive to the supersaturation (classical theory of nucleation, for example, [42,43]), as is the precipitated $\phi$ of the γ'-precipitates. The larger values $N_a^{'}$ and smaller precipitate dimensions (Table 2) observed for the higher-supersaturated Ni-Al-Cr-W alloy results from a larger nucleation current coupled with slower growth and coarsening kinetics [21]. For both alloys, $N_a^{'}$ decreases with increasing aging time as their dimensions increase, indicating that nucleation ends prior to 4 h. Consistent with the constant elemental partitioning, the $A_a^{'}$ values for both alloys are nearly constant as a function of aging time (Table 2). The Ni-Al-Cr-W alloy has a larger $A_a^{'}$ (~ 38%) value than the Ni-Al-Cr alloy (~ 25%), which is a direct result of the larger matrix supersaturation.

Figures 5 and 6 are wide field-of-view images of the nanostructure, after the onset of the sphere-to-cube morphological transition. Although the grains in the Ni-Al-Cr alloy exhibit a random texture with respect to the ingot length (Figure 1), the orientation of the particular grains imaged in Figure 5 may be inferred from the observed nanostructure. For the 64 h Ni-Al-Cr specimen (Figure 5a), the majority of imaged γ'-precipitates are elongated slightly and uniformly in one direction, while the γ'-precipitates for the 264 h specimen (Figure 5b) are mostly equiaxed, which suggests a <110>-type and <001>-type textures with respect to the surface normal, respectively. In the 64 h aging state, the cuboidal γ'-precipitates (exhibiting significantly rounded corners) align in long strings along an <001>-type direction. The imaged strings consist of as many as 16 aligned γ'-precipitates and exhibit a slight curvature along the string extension. At 264 h, the alignment changes and the highly faceted-cuboids (with a great degree of corner matching of adjacent γ'-precipitates) appear to align in clustered groups along the orthogonal <001>-type directions, which permits precipitate free areas within the imaging plane. Compared to the Ni-Al-Cr specimens, the γ'-precipitates in the 64 h aged Ni-Al-Cr W specimens (Figure 6) are highly faceted and have a smaller $< \lambda_{e-e}^{PS} >$ value of 23 ± 3 nm (Table 2). The dense packing for this aging state prevents the γ'-precipitates from aligning in long strings. With further aging to 264 h, although $N_v$ decreases, the change in $< \lambda_{e-e}^{PS} >$ to 24 ± 5 nm is extremely slight, suggesting the γ'-precipitates undergo some alignment along the elastically soft <001>-type directions, which is most likely due to an increase in the long-range elastic stress field of the γ'-precipitates with increasing $<R>$ and the less dense spatial distribution (Figure 6b). The alignment of γ'-precipitates is coupled with a decrease in microhardness after a constant value between 4 to 64 h of aging (Figure 2b).

### 3.3. Stereological analysis

When the polish relief depth is comparable to the feature size, projection effects need to be considered. Given that the γ'-precipitates range in size from 10 to 100 nm, the polish relief depths (Table 2) necessary to obtain sufficient imaging contrast between the γ-phase and γ'-phase are non-negligible, and need to be treated explicitly with stereological corrections. A surface with significant polish relief relative to the γ'-precipitate dimensions is analogous to a finite section with one transparent phase and one opaque, as in conventional dark-field TEM imaging.

For a finite section, the $\phi$, $N_v$, and the interfacial area per unit volume ($S_v$) values are determined from the following stereological relationships [37]:

$$\phi = A'_a - h \cdot \left( \frac{1}{2} P'_l - N'_a \cdot \frac{h \cdot <H'>}{h + <H'>} \right); \qquad (2)$$

$$N_v = \frac{N'_a}{h + 2 <R>}; \qquad (3)$$

and

$$S_v = 2P'_l - 4N'_a \cdot \frac{h \cdot <H'>}{h + <H'>}; \qquad (4)$$

where $A'_a$ is the precipitated areal fraction; $h$ is the polish relief depth; $P'_l$ is the number of points per unit length; $N'_a$ is the areal density of precipitates; $<H'>$ is the mean caliper measurement; and $<R>$ is the mean precipitate radius in three-dimensions. The measurements on the polish relief surfaces are displayed in Table 2, while the properties calculated from Equations 2-4 are presented in Table 3. To obtain a measurement of $<R>$ (Table 3), the $PSD_{PS}$ of the $<R_{PS}>$ values are converted to three-dimensional PSDs employing methods outlined in reference [44]. This conversion procedure yields values for $<R>$ at $t = 264$ h in agreement with the measurements made on TEM micrographs reported in [21].

Figure 7 demonstrates that the stereologically calculated $\phi$ values (Equation 2) are constant within experimental error for the times considered and that these values agree with the phase extraction measurements of $\phi^{eq}$ at 264 h. Since these phase extraction measurements agree closely with THERMO-CALC predictions of $\phi^{eq}$, the alloys appear to be at their $\phi^{eq}$ values for all aging times investigated. The exception being the calculated $\phi$ value for the 4 h Ni-Al-Cr specimen is 14 ± 5%, which is somewhat smaller than the $\phi^{eq}$ value of 18.9 %. This particular measurement may be artificially low due to precipitates falling-out during etching, as $h$ is relatively large when compared to <$R$>. In general, the agreement between these independent measurements demonstrates that the application of stereological procedure to both alloys is reasonable. Finally, since the transformation has reached $\phi^{eq}$ and the two-phase microstructure is chemically invariant, diffusion-limited growth has finished and the alloys are within the coarsening regime. If the microstructures are also shown to be time-invariant, then the coarsening is considered to be in a stationary-state, which is rarely achieved [45] and considered in more detail below.

### 3.4. Coarsening kinetics

During stationary-state coarsening <$R$>, increases as:

$$< R(t) >^3 - < R(0) >^3 = Kt ; \qquad (5)$$

while the mean interfacial per unit volume <$S_v$> decreases as [46]:

$$< S_v(t) >^{-3} - < S_v(0) >^{-3} = K_S t ; \qquad (6)$$

where $K$ and $K_S$ are the coarsening rate constants and <$R(0)$> and <$S_v(0)$> are the pertinent values at the onset of coarsening, which is not the zero time value. For the finite sections under consideration, the *calculated* <$R$> values are confounded by the 2D-to-3D conversion, as well as by the areal approximation of circular equivalence through the spheroidal-to-cuboidal transition, while the measurement of $S_v$ can be made without assuming a specific precipitate morphology [37]. The quantity $S_v$ provides, however, no information about the morphology or size distribution of individual precipitates. Based on Equations 5 and 6, the constants $K$ and $K_s$ are obtained from the slopes of the corresponding plots in Figures 8 and 9, where the error is determined from a linear regression analysis of the best-fit of the experimental data. For both $<S_v>^{-3}$ and $<R>^3$, a linear relationship with time obtains, and the Ni-Al-Cr alloy coarsens at rate that is 3.3 times faster than the tungsten containing quaternary alloy. During stationary-state coarsening, independent of the number of components, $N_v$ is found to be proportional to $t^{-1}$ [47] based on a generalized Lifshitz-Slyozov-Wagner coarsening model. The experimental power-law time dependencies (Figures 10), of $-0.84 \pm 0.04$ for the Ni-Al-Cr alloy and $-0.88 \pm 0.07$ for the Ni-Al-Cr-W alloy, deviate from the stationary-state prediction, which is evidence that the systems have not achieved a true stationary-state and are most likely in a quasi-stationary state.

### 3.5. Precipitate-size distributions (PSDs), pair-correlation functions (PCFs) and edge-to-edge interprecipitate distance distributions (IDDs).

Quasi-stationary state coarsening behavior and the elastically-induced alignment of the γ'-precipitates in these two alloys can be better understood by the temporal evolution of the PSDs, PCFs, and IDDs). Following [12], the area-equivalent radius of a circle, $R_{PS}$, which yields one effective dimension measure, is used to determine the $PSD_{PS}$. This nomenclature distinguishes $PSD_{PS}$ from PSD, which differs because it is determined from measured $R_{PS}$ values rather than measured $R$ values, and denotes that the measured quantities are taken from a planar section (PS); in this case, cross-sectional SEM images of the 3D-nanostructure. Figures 11 and 12 present the temporal evolution of the scaled $PSD_{PS}$ for the Ni-Al-Cr and Ni-Al-Cr-W alloys, which are compared to the $\phi^{eq} = 0.2$ and $\phi^{eq} = 0.3$ stationary-state predictions of Akaiwa and Voorhees [27] *for coarsening driven by a reduction in interfacial free energy (Ostwald ripening) in systems with spherical (stress-free) precipitate geometries*. Note that the stationary-state predictions are for volume fractions that are reasonably close to the measured $\phi^{eq}$s of 0.189 and 0.308 in the Ni-Al-Cr and Ni-Al-Cr-W alloys, respectively. Figures 11 and 12 demonstrate that the experimental $PSD_{PS}$s evolve temporally and are not self-similar indicating that system has not achieved a stationary-state.

For the 4 h Ni-Al-Cr specimens (Figure 11), the $PSD_{PS}$ is broader than and with a similar height maximum to the stationary-state prediction yielding an extended tail for large values of $R_{PS}/<R_{PS}>$. The maximum appears at a smaller $R_{PS}/<R_{PS}>$ value of 0.9 than the stationary-state prediction of 1.15, according to the AV model. With further aging to 64 h, the $PSD_{PS}$s become narrower as the height of the maximum increases and remains at $R_{PS}/<R_{PS}> = 0.9$. In the final aging state, the maximum value shifts to the $R_{PS}/<R_{PS}>$ value predicted by the AV model [27], and the distribution matches more closely the stationary-state prediction than at shorter aging times. Similarly, analysis of the 4 h Ni-Al-Cr-W specimens (Figure 12) results in $PSD_{PS}$ with a maximum at a smaller $R_{PS}/<R_{PS}>$ value, 0.9, than predicted; however, the maximum is sharper than predicted and than observed for the Ni-Al-Cr alloy. With further aging to 64 h, $PSD_{PS}$s remain skewed towards smaller scaled radii with its maximum at $R_{PS}/<R_{PS}> = 0.9$. Rather than increasing, which is observed for Ni-Al-Cr alloy, the height of this maximum decreases for these aging times. For the final aging state, the $PSD_{PS}$ approaches a stationary-state distribution but does not match exactly the AV stationary-state prediction, as the values at smaller $R_{PS}/<R_{PS}>$ values are still greater than predicted.

The temporal evolution of the normalized $PCF_{PS}$ for both alloys is displayed in Figure 13. In all the $PCF_{PS}$, an exclusion zone, typically $2<R_{PS}>$, exists, and it is the region around a γ-precipitate center that is nearly precipitate free. Despite the elastically driven changes in the Ni-Al-Cr spatial distributions with aging, the normalized $PCF_{PS}$ values as a function of radial distance outside the exclusion zone do not deviate strongly from unity and the $PCF_{PS}$ are approximately time-invariant (Figure 13 a). For these profiles, the average probability of finding a precipitate center at $2.75R_{PS}$ is only marginally greater than at other distances outside the exclusion zone. The $PCF_{PS}$ values for the Ni-Al-Cr-W alloy deviate strongly from unity, due to their more dense packing (Figure 13b). A sharp peak in the $PCF_{PS}$ at $r/<R_{PS}> = 2.75$, which has a normalized height of ~1.4, is observed for $t \le 64$.

Adjacent to the sharp peak, a slight minimum, with a normalized height of 0.9, at $r/\langle R_{PS}\rangle = 4.25$, reflects a smaller probability of finding γ'-precipitates at this distance and is considered to be a measure of the extension of the center-to-center clustering zone. Interestingly, with the onset of elastically-induced alignment for the final aging state, the height of the clustering peak decreases significantly to 1.2, no minimum is detected, and the profile flattens.

The frequency distributions of $\lambda_{e-e}^{PS}$ (divided into a constant number of intervals over the full range of values) and their $\text{IDD}_{PS}$s as function of aging time are displayed in Figures 14 and 15. Consistent with less dense γ'-packing in the Ni-Al-Cr alloy, the γ'-precipitates are more widely spaced than those in the Ni-Al-Cr-W alloy with a range of $\lambda_{e-e}^{PS}$ values that is approximately double for a given aging time. For aging times up to 64 h, the frequency distributions (Figure 14 a) for the Ni-Al-Cr alloy become narrower, while the maximum in the corresponding $\text{IDD}_{PS}$ shifts from 0.8 to 0.2 $\lambda_{e-e}/\langle R\rangle_{PS}$ and increases in intensity from 1.0 to 1.5. These observed changes in these $\lambda_{e-e}^{PS}$ distributions are incremental with successive time steps and reflect the influence of elastic interactions on the uniformly distributed γ'-precipitates ($t = 4$ h) to aligned <001>-oriented strings ($t = 64$ h) over this period of time. At 264 h, the cuboidal γ'-precipitates align in clustered groups, the $\text{IDD}_{PS}$ maximum shifts towards a larger $\lambda_{e-e}/\langle R\rangle_{PS}$ value of 0.65 as its height decreases to 1.3. Despite a complex morphological evolution, the Ni-Al-Cr-W does not exhibit such pronounced changes. The $\lambda_{e-e}^{PS}$ frequency distributions and $\text{IDD}_{PS}$s displayed in Figure 15, where $\text{IDD}_{PS}$s are approximately time-invariant. The frequency distribution for the 264 h aging state, however, is slightly skewed towards smaller values such that the 264 h $\langle \lambda_{e-e}^{PS}\rangle$ of 24 ± 5 nm (Table 2) is nearly equivalent to the 64 h value of 23 ± 3 nm. This result is consistent with the elastically-induced alignment in the soft <001>-type direction observed for the 264 h aging state.

IV. Discussion

### 4.1. Morphological evolution

At early aging times, the morphology of γ'-precipitates is spheroidal. With further aging time and increasing <R>, the γ'-precipitates become less rounded and {100}-type facets form as a cuboidal morphology develops. The <R> value at which the microstructure becomes cuboidal is estimated to be ≈ 88 nm ($t = 64$ h) for the Ni-Al-Cr alloy and ≈ 50 nm ($t ≈ 32$ h) for the Ni-Al-Cr-W alloy based on the micrographs in Figure 3. The spheroid-to-cuboid transition can be understood as a competition between the elastic self-energy and precipitate/matrix interfacial free energy [48], which ultimately determines the equilibrium shape. As a γ'-precipitate grows, the elastic self-energy increases as $R^3$, while the interfacial energy of each precipitate increases as $R^2$. Hence, as growth and coarsening proceeds in elastically stressed systems the elastic energy becomes more influential in determining precipitate morphology. In a model developed by Thompson et al. [49], the relative magnitude of the elastic energy to the interfacial energy, $σ^{αβ}$, is quantified by a dimensionless parameter, $L$:

$$L = \frac{ε^2 C_{44}}{σ^{γγ'}} R; \qquad (7)$$

where $ε$ is the unconstrained lattice parameter misfit strain, defined as $ε = (a^{γ'} - a^γ)/a^γ$ (where $a^γ$ is the temperature-dependent lattice parameter of phase γ) and $C_{44}$ is an elastic constant for the γ-matrix phase. For a precipitate with a purely dilatational misfit in a Ni matrix, $L = 2$ to $4$ for the equilibrium shape of a 4-fold symmetric cuboid [49]. Substituting values for the Ni-Al-Cr alloy of $ε = 0.0022 ±$ 0.0007 [50], $σ^{γγ'} = 0.023 ± 0.007$ J m$^{-2}$ [51], $C_{44} = 95 ± 10$ GPa [52], and $R = 88$ nm, yields the calculated value of $L$ to be $1.8 ± 1.4$, which is in close agreement with the model taking into account the experimental uncertainties of all the parameters employed in Eqn. (6). That the transition from spheroids-to-cuboids occurs at a smaller $R$ value of 50 nm in the Ni-Al-Cr-W alloy most likely results from the larger lattice parameter misfit between the γ- and γ'-phases, since $L$ is directly proportional to $ε^2$. Assuming to first order that $C_{44}$ and $σ^{γγ'}$ are the same as for the Ni-Al-Cr alloy, the $\varepsilon$ value for the Ni-Al-Cr-W alloy is estimated to be 0.0029 ± 0.0009 by equating L for the two $R$ values at the morphological transition.

The γ'-precipitate alignment results from the minimization of the elastic interactions among the γ'-precipitates, where the interaction energy depends on the elastic anisotropy, as well as the difference in elastic constants between the phases, and the magnitude of $\varepsilon$ [1,53]. The interaction of long-range elastic fields with one another gives rise to a configurational force that is attractive at long distances and repulsive at short distances [54,55]. In the Ni-Al-Cr alloy, near the spheroid-to-cuboid transition, the γ'-precipitates, which have a significant degree of rounding, appear to align initially in isolated strings of precipitates along one <001>-type direction. As γ'-precipitates increase in size and become more faceted, they align in orthogonal <001>-type directions in clustered groups of γ'-precipitates. Voorhees and Johnson [55] demonstrate that if the negative elastic energy along the <100>-directions and the positive elastic energy along the <110>-directions have approximately the same magnitude, then the most favorable three-dimensional spatial alignment for misfitting cuboidal precipitates is an isolated one-dimensional string, suggesting that this is true for the microstructure observed for the 64 h aged Ni-Al-Cr specimens. Further investigation, however, is needed to ascertain how the relationship between the evolving morphologies and the magnitude of elastic energies along the <100>- and <110>-type directions evolve. In the Ni-Al-Cr-W alloy, a similar microstructure of clustered groups of γ'-precipitates is observed for both the spheroidal and cuboidal morphologies of γ'-precipitates over the entire time scale investigated. As shown in Figure 13 b, these cluster zones of γ'-precipitates extend 4.25 <$R_{PS}$>. Unlike the Ni-Al-Cr alloy, isolated strings of γ'-precipitates are not observed in the Ni-Al-Cr-W alloy (Figure 2 (e-h) and Figure 5) and are not possible, even at shorter times, as inter-string interactions at the higher volume fraction (Figure 12 b) prevent their formation [56]. Recently, Lund and Voorhees [56] characterized in three-dimensions, on a micron scale, the γ'-microstructure of a Ni-24.0 Co-5.0 Cr-2.5 Mo-4.0 Al-4.0 Ti wt. % alloy for $\phi = 0.272$. In the Lund-Voorhees alloy, with a similar $\phi$ value to the alloys investigated, the γ'- precipitates align in isolated sheets. A two-dimensional slice within an isolated sheet of γ'-precipitates yields an array of two-dimensional clusters of γ'-precipitates with a limited extension [56], consistent with the experimental observations for the later aging times presented herein.

### 4.2. Coarsening kinetics

Recent work on γ'-coarsening at 1173 K [56] in Ni-Al alloys shows that $<S_v>$ follows a $t^{-1/3}$ law, even at longer aging times ($\approx$ 40 h) when the microstructure consists of rod-shaped precipitates and the elastic energy contribution to the total energy of system is significant; a nonlinearity in $<R^3>$ is, however, observed and attributed to nonequiaxed precipitate morphologies. The microstructures in the systems under investigation here are equiaxed and not as coarse. Since the magnitude of the elastic energy effects scale as $R^3$, these effects are most likely not as large as for the Ni-Al alloys presented in [56]. Therefore, the Ni-Al-Cr and Ni-Al-Cr-W alloys we study are expected to be governed by the reduction in interfacial free energy. Figures 8 and 9 demonstrate that the coarsening kinetics of *both* $<R>$ and $<S_v>$ obey the $t^{1/3}$ and $t^{-1/3}$, respectively. The measured temporal power-law dependencies of $N_v$ from the stereological analysis of the two-dimensional structure are –0.84 ± 0.04 for the Ni-Al-Cr alloy and –0.88 ± 0.07 for the Ni-Al-Cr-W alloy (Figure 10), which are in good agreement with one another, but deviate from the stationary-state prediction of –1; therefore, the alloys have not truly achieved a stationary-state. The PSD$_{PS}$s (Figures 11 and 12) for both alloys evolve continuously with time verifying that they have not yet achieved a stationary-state; however, at the longest aging time, 264 h, the PSD$_{PS}$s for both alloys are similar to but not exactly the stationary-state prediction [27].

The coarsening kinetics is significantly affected by the addition of W. Although the precipitated $\phi$ value is smaller, the Ni-Al-Cr alloy under investigation coarsens at a rate ($K$) that is 3.3 times faster than the Ni-Al-Cr-W alloy. This behavior suggests that chemical composition plays a critical role in the coarsening behavior of these systems. For the Kuehmann-Voorhees coarsening model of a ternary alloy in stress-free systems [28], the coarsening rate constant is a function of: (i) the magnitude of the partitioning between the phases ($p_i = C_i^{\gamma',eq} - C_i^{\gamma,eq}$); (ii) the relative ratios of curvatures of the Gibbs free-energy surfaces for the matrix phase ($G_{,ij}^k$); and (iii) the solute diffusivities ($D_{ii}$) of component $i$, where the ratio $p_i / D_{ii}$ is critical in determining $G_{,ij}^k$ contributions. Multicomponent coarsening models assume that the off-diagonal terms of diffusivity matrix (ergo these terms are zero in the susceptibility and Onsager matrices) and the chemical potential is everywhere zero, which are strong assumptions, as it has been shown recently that the details of the diffusion mechanism in a Ni-Al-Cr alloy aged at 873 K strongly affect the kinetic pathway of precipitation [57]. As found by APT measurements [21], the addition of W influences the elemental partitioning in the following manner: (i) Cr partitions more strongly to the γ-matrix, by a factor of 1.80 ± 0.09 than in the Ni-Al-Cr alloy (as determined from equilibrium partitioning ratios, $C_i^{\gamma',eq} / C_i^{\gamma,eq}$, that we measured for both alloys); (ii) Al partitions more strongly to γ'-precipitates, by a factor of 1.31 ± 0.05; and, (iii) there are approximately twice as many W atoms per unit volume in the γ'-precipitates than in the γ-matrix. The stronger phase partitioning of Al and Cr for the Ni-Al-Cr-W alloy is also reflected in the $p_{Al}$ and $p_{Cr}$ values calculated from the ICP measurements (Table 1), where their absolute values increase by 1.64 at.% and 3.69 at.% for Al and Cr, respectively. For the Ni-Al-Cr-W alloy, the value of $p_W$ of 1.46 ± 0.03 is an order of magnitude smaller than $p_{Al}$ of 10.62 ± 0.18, so the $p_W$ influence on the coarsening constant may be small. Relative to its combined influence on elemental partitioning ($p_i$), the effect of W on the solute diffusivities is minor, only reducing the Al and Cr tracer diffusivities in the γ-phase by 2-3 % (Table 1) [32,33]. The tracer diffusivity of W in the γ-matrix at 1073 K, 4.93 x 10$^{-18}$ m$^2$ s$^{-1}$ [32,33], is approximately one order of magnitude smaller than the tracer diffusivity of Cr in the γ-matrix, the next slowest diffusing species, and counters the effect of a small value of $p_W$. Given that the Al and Cr tracer diffusivities are similar for the alloys studied, it is likely the influence of W on the partitioning behavior of the elemental species, as well as its small mobility, causes the coarsening rate in the Ni-Al-Cr-W alloy to be significantly slower than in the Ni-Al-Cr alloy. A similar coarsening behavior has been established for Ni-Al-Mo alloys [12,13], where the diffusivity of Mo in Ni is smaller than that of Al in Mo.

## V. Summary and Conclusions

The grain structure, hardness, precipitate morphology, precipitate spatial correlations, and coarsening kinetics are explored in detail for two $\gamma/\gamma'$ alloys, Ni-9.8 Al-8.3 Cr (at.%) and Ni-9.7 Al-8.5 Cr-2.0 W (at.%), isothermally aged at 1073 K from 4 h to 264 h, where the unconstrained lattice parameter misfits are estimated to be $0.0022 \pm 0.0007$ and $0.0029 \pm 0.0009$, respectively, leading to the following results:

- The Ni-Al-Cr alloy exhibits hardness values that vary between 2.15-2.5 GPa (Figure 2). Its grain structure is fairly coarse, equiaxed, and twinned with different crystallographic orientations (Figure 1). By adding 2 at.% W, the microhardness increases by 40-50 % (Figure 2), the grains are less coarse, elongated radially, and exhibit a <110>–texture (Figure 1).

- Employing stereological corrections for a section with a finite thickness (Equations 2-4), the volume fraction ($\phi$), mean precipitate radii ($<R>$), interfacial area per unit volume ($S_v$), and number density per unit volume ($N_v$) of the $\gamma'(L1_2)$-precipitates protruding from a chemically etched $\gamma$(fcc)-matrix are precisely characterized with SEM and AFM measurements of the polish relief depth (Figure 4, Table 2). For the entire aging sequence, the value of $\phi$ is constant and is at its equilibrium value (Figure 7). The coarsening kinetics of the quantities $<R>$ and $S_v$ obey $t^{1/3}$ and $t^{-1/3}$ laws (Figures 8 and 9), establishing that coarsening is dominated by a reduction in interfacial energy. The $N_v$ power law dependencies of $-0.84 \pm 0.04$ and $-0.88 \pm 0.07$ for the Ni-Al-Cr and Ni-Al-Cr-W alloys, respectively (Figures 10), deviate from the model stationary-state prediction of $-1$, suggesting the systems are still undergoing quasi-stationary state coarsening. That is, they have not achieved a stationary state.

- Initially, the nanostructure in the Ni-Al-Cr alloy evolves temporally from isolated spheroidal γ'-precipitates uniformly distributed, to strings of cuboidal precipitates aligned along one <100>-type direction, and finally, to highly-faceted cuboids aligned in the orthogonal <100>-type directions into clustered groups (Figure 3 (a-d) and Figure 5). Despite this, the two-dimensional pair correlation functions are time invariant (Figure 13 a). However, the two-dimensional interprecipitate distance distributions ($IDD_{PS}$s) of edge-to-edge distances (Figure 14) evolve temporally. With string alignment, these $IDD_{PS}$s narrow, become more intense, and shift towards smaller normalized distances (Figure 14 b). As the γ'-precipitates cluster in groups, a broader and flatter $IDD_{PS}$ that shifts toward larger distances results.

- Due to a stronger nucleation current from a larger matrix supersaturation coupled with slower coarsening kinetics, the Ni-Al-Cr-W alloy contains a larger value of $N_v$ of smaller γ'-precipitates that are more densely packed than in the Ni-Al-Cr alloy (Figure 3). For the shorter aging times (4 and 16 h), the γ'-precipitates are spheroidal and some coagulation and coalescence among the γ'-precipitates leads to irregular morphologies. With further aging, the γ'-precipitates become cuboidal and about 5% are nonequiaxed (Figure 6). Through the morphological transition (~32 h) up to 64 h of aging, a sharp peak in the two-dimensional pair-correlation functions (Figure 13 b) is time invariant and attributed to this dense packing. For the final aging state, a less dense distribution with larger long-range interactions allows γ'-precipitates to align elastically along the soft <100>-type directions into groups and results in a flatter pair-correlation function. The edge-to-edge $IDD_{PS}$s are approximately but not exactly time-invariant (Figure 15).

- Compared to stationary-state predictions for stress-free coarsening at nonnegligible volume fractions [27], the two-dimensional precipitate size distributions (Figures 11 and 12) are initially skewed towards smaller radii. With increasing aging time, these $PSD_{PS}$s approach the predicted distribution and are in near agreement for the final aging time (264 h).

- The addition of 2 at.% W to the Ni-Al-Cr alloy reduces the coarsening rate constant by more than one-third. Tungsten also significantly influences the elemental partitioning of the other constituents to the γ- and γ'-phases (Table 1). It is likely that tungsten's influence on the partitioning, as well as its inherently small diffusivity in Ni-Al-Cr, are responsible for the observed deceleration of the coarsening kinetics.

**Acknowledgements.** This research was supported by the National Science Foundation (NSF), Division of Materials Research, under contract DMR-0241928. CKS received partial support from NSF graduate research fellowships. TDZ received partial support from Northwestern University's Walter P. Murphy undergraduate research grant and an NSF Research Experiences for Undergraduates (REU) grant. The authors would like to thank Dr. C. E. Campbell for diffusivity calculations, Prof. P. W. Voorhees for model predictions of the precipitate size distributions, Mr. Ben Myers for assistance with SEM imaging, Mr. Richard Karnesky for a program and aid in determining two-dimensional edge-to-edge spacings [40] and NSF REU students Ms. Gillian Hsieh, Mr. Luis de la Cruz, Mr. A. Vaynman, and Mr. N. Disabato for assistance in specimen preparation and characterization.. Drs. David Rowenhorst, Roberto Mendoza, and Alan C. Lund are graciously thanked for fruitful discussions.

*List of Symbols*

| | |
|---|---|
| $a^{\gamma}$ | lattice parameter of phase γ |
| $A_a^{'}$ | precipitated areal fraction for a section with finite thickness |
| $C_i^o$ | overall alloy concentration of component $i$ |
| $C_i^{\gamma,eq}$ | equilibrium concentration of component $i$ in phase γ |
| $\boldsymbol{D}_{ii}$ | diffusivity matrix |
| $f$ | percentage of γ'-precipitates whose aspects ratios are greater than 1.75 |
| $G_{,ij}^k$ | partial derivatives of the molar Gibbs free-energy of phase $k$ |
| $h$ | etching height |
| $<H'>$ | mean caliper distance |
| $K^{KV}$ | coarsening rate constant for $R$ according to the KV model |
| $N_a^{'}$ | number of precipitates per unit volume (areal density) for finite section |
| $N_v$ | number of precipitates per unit volume (number density) |
| $p_i$ | difference in elemental partitioning in equilibrium, $C_i^{\gamma',eq} - C_i^{\gamma,eq}$ |
| $P_l^{'}$ | points per unit length for a finite section |
| $r$ | radial distance from precipitate center |
| $R_{PS}$ | circular-equivalent radius of an intersected precipitate area in a planar section |
| $R$ | precipitate radius |
| $<R(t)>$ | time-dependent mean precipitate radius |
| $<R(0)>$ | mean precipitate radius at the onset of coarsening, which is not $t = 0$ |
| $S_v$ | interfacial area per unit volume |
| $t$ | aging time |
| $V_m$ | molar volume of the precipitate |
| | |
| $\varepsilon$ | unconstrained lattice parameter misfit strain |
| $\phi$ | precipitated volume fraction |
| $\phi^{eq}$ | equilibrium precipitated volume fraction |
| $\gamma$ | matrix phase |
| $\gamma'$ | Ni$_3$(Al,Cr)-type precipitated phase |
| $\lambda_{e-e}$ | edge-to-edge interprecipitate distance |
| $\lambda_{e-e}^{PS}$ | edge-to-edge interprecipitate distance in a planar section |
| $<\lambda_{e-e}^{PS}>$ | mean edge-to-edge interprecipitate distance in a planar section |
| $\sigma$ | standard deviation |
| $\sigma^{\gamma\gamma'}$ | interfacial free energy between $\gamma$ and $\gamma'$ phases |
| | |
| APT | Atom probe tomography |
| AFM | Atomic force microscopy |
| FIM | Field ion microscopy |
| IDD | Interprecipitate distance distribution |
| KV | Kuehmann-Voorhees model |
| PCF | Pair-correlation function |
| PSD | Particle size distribution |
| SEM | Scanning electron microscopy |

CK Sudbrack, TD Ziebell, RD Noebe, DN Seidman, "*Effects of a tungsten addition on the morphological evolution, spatial correlations, and temporal evolution of a model Ni-Al-Cr superalloy*" Acta Mater. (2007), in press.

CK Sudbrack, TD Ziebell, RD Noebe, DN Seidman, "*Effects of a tungsten addition on the morphological evolution, spatial correlations, and temporal evolution of a model Ni-Al-Cr superalloy*" Acta Mater. (2007), in press.

Table 1. Inductively coupled plasma (ICP) atomic emission spectroscopy measurements[a] of the bulk and equilibrium phase compositions and the tracer diffusivities calculated for the two alloys investigated[b,c]

| Alloy | Property | Ni | Al | Cr | W |
|-------|----------|-----|-----|-----|-----|
| Ni-Al-Cr | Average bulk composition as measured with ICP atomic emission spectroscopy (at.%) | 81.90 ± 0.13 | 9.76 ± 0.10 | 8.34± 0.08 | – |
| | Equilibrium composition of the γ'-precipitates, $C_i^{\gamma',eq}$, as measured with ICP atomic emission spectroscopy (at.%) | 76.60 ± 0.18 | 17.41 ± 0.17 | 5.99± 0.06 | – |
| | Equilibrium composition of the γ-matrix, $C_i^{\gamma,eq}$, as measured with ICP atomic emission spectroscopy (at.%) | 82.71 ± 0.12 | 8.43± 0.08 | 8.86 ± 0.09 | – |
| | Magnitude of the elemental partitioning, $p_i = C_i^{\gamma',eq} - C_i^{\gamma,eq}$ (at.%) | -6.11 ± 0.22 | 8.98 ± 0.19 | -2.87 ± 0.11 | |
| | Tracer diffusivity at 1073 K (x $10^{-18}$ m$^2$ s$^{-1}$), $D_i$ | 6.06 | 13.9 | 5.13 | – |
| | $D_i / D_{Ni}$ | 1 | 2.29 | 0.85 | – |
| Ni-Al-Cr-W | Average bulk composition as measured with ICP atomic emission spectroscopy (at.%) | 79.81 ± 0.13 | 9.74 ± 0.10 | 8.49 ± 0.08 | 1.96 ± 0.02 |
| | Equilibrium composition of the γ'-precipitates, $C_i^{\gamma',eq}$, as measured with ICP atomic emission spectroscopy (at.%) | 76.21 ± 0.18 | 16.85 ± 0.17 | 3.94 ± 0.04 | 3.00 ± 0.03 |
| | Equilibrium composition of the γ-matrix, $C_i^{\gamma,eq}$, as measured with ICP atomic emission spectroscopy (at.%) | 81.75 ± 0.12 | 6.23 ± 0.06 | 10.48 ± 0.10 | 1.54 ± 0.02 |
| | Magnitude of the elemental partitioning, $p_i = C_i^{\gamma',eq} - C_i^{\gamma,eq}$ (at.%) | 5.54 ± 0.21 | 10.62 ± 0.18 | -6.56 ± 0.11 | 1.46 ± 0.03 |
| | Tracer diffusivity at 1073 K (x $10^{-18}$ m$^2$ s$^{-1}$), $D_i$ | 4.92 | 13.5 | 5.05 | 0.493 |
| | $D_i / D_{Ni}$ | 1 | 2.74 | 1.03 | 0.10 |

[a.] Compositions measured by ICP atomic-emission spectroscopy are assumed to have a 1% standard error for the concentrations of the solute species, where standard error propagation methods [41] are employed to determine the error in Ni concentration.

[b.] Equilibrium composition of the γ- and γ'-phases are measured directly with ICP atomic-emission spectroscopy from electrolytically separated phases of specimen sections aged for 264 h at 1073 K.

[c.] Tracer diffusivities [32] for face-centered cubic solid solution are calculated with the thermodynamic database developed by Saunders [31] and the mobility database developed by Campbell [33] for Ni-based superalloys.

Table 2. Nanostructural measurements[a,b] obtained directly from SEM and AFM images of chemically etched Ni-9.8 Al-8.3 Cr at.% and Ni-9.7 Al-8.5 Cr-2 W at.% samples

| Alloy | $t$ (h) | $h \pm \sigma$ [c] (nm) | $<H'> \pm \sigma$ [d] (nm) | $<\lambda_{e-e}>_{PS} \pm \sigma$ [d] (%) | $A'_a \pm \sigma$ [d] (%) | $N'_a \pm \sigma$ [d] x $10^{13}$ (m$^{-2}$) | $P'_l \pm \sigma$ [d] (µm$^{-1}$) | $f \pm \sigma$ [d] (%) |
|---|---|---|---|---|---|---|---|---|
| NiAlCr | 4 | $20 \pm 4$ | $41 \pm 3$ | $20 \pm 3$ | $24 \pm 4$ | $15.1 \pm 0.5$ | $14.3 \pm 0.6$ | $1.61 \pm 0.05$ |
| | 16 | $17.0 \pm 1.7$ | $75 \pm 6$ | $26 \pm 6$ | $24.7 \pm 1.2$ | $5.06 \pm 0.09$ | $7.29 \pm 0.17$ | $2.14 \pm 0.04$ |
| | 64 | $35 \pm 4$ | $119 \pm 7$ | $31 \pm 7$ | $27 \pm 3$ | $2.56 \pm 0.05$ | $5.98 \pm 0.14$ | $0.654 \pm 0.012$ |
| | 264 | $25.8 \pm 2.8$ | $384 \pm 10$ | $62 \pm 10$ | $22.7 \pm 1.2$ | $1.166 \pm 0.027$ | $3.26 \pm 0.08$ | $2.75 \pm 0.07$ |
| NiAlCrW | 4 | $3.8 \pm 0.7$ | $37.0 \pm 2.0$ | $9.1 \pm 2.0$ | $39 \pm 4$ | $32.3 \pm 0.6$ | $23.7 \pm 0.5$ | $2.35 \pm 0.05$ |
| | 16 | $9.2 \pm 1.4$ | $56.6 \pm 2.5$ | $13.2 \pm 2.5$ | $40 \pm 4$ | $14.0 \pm 0.3$ | $17.2 \pm 0.4$ | $4.91 \pm 0.12$ |
| | 64 | $7.1 \pm 0.7$ | $79 \pm 3$ | $23 \pm 3$ | $36 \pm 3$ | $6.41 \pm 0.16$ | $11.00 \pm 0.25$ | $4.55 \pm 0.11$ |
| | 264 | $26 \pm 3$ | $118 \pm 5$ | $24 \pm 5$ | $37.9 \pm 1.6$ | $3.28 \pm 0.07$ | $8.00 \pm 0.18$ | $5.28 \pm 0.12$ |

[a.] $h$ is the polish relief depth as measured from line-traces across AFM images

[b.] $<H'>$ is the mean caliper length; $<\lambda_{e-e}>_{PS}$ is the edge-to-edge interprecipitate distance; $A'_a$ is the areal fraction; $P'_l$ is the points per unit length; $N'_a$ is the areal density; $f$ is the percentage of precipitates with aspect ratios greater than 1.75; of the γ'-precipitates as measured directly from SEM images of 1000 to 3500 γ'-precipitates for a given aging time.

[c.] Standard deviation of multiple independent AFM measurements

[d.] Stand error propagation [41] based on counting errors and tracing errors

Table 3. Temporal evolution of the nanostructural properties[a] of the γ'-precipitates in Ni-9.8 Al-8.3 Cr at.% and Ni-9.7 Al-8.5 Cr-2 W at.%  aged isothermally at 1073 K

| Alloy | $t$ (h) | $N_{ppt}$[b] | $\langle R \rangle_{PS} \pm \sigma$ (nm) | $\langle R \rangle \pm \sigma$ (nm) | $N_v \pm 2\sigma$ x $10^{20}$ ($m^{-3}$) | $\phi \pm \sigma$ (%) | $S_v \pm \sigma$ x $10^6$ ($m^{-1}$) |
|---|---|---|---|---|---|---|---|
| Ni-Al-Cr | 4 | 869 | 20.8 ± 1.8 | 31.1 ± 2.7 | 29 ± 5 | 14 ± 5 | 20 ± 8 |
| | 16 | 3179 | 37.3 ± 2.9 | 58 ± 5 | 6.8 ± 1.1 | 19.7 ± 1.3 | 12 ± 4 |
| | 64 | 2752 | 59 ± 4 | 88 ± 5 | 2.1 ± 0.5 | 19 ± 3 | 9.2 ± 2.8 |
| | 264 | 1418 | 81 ± 6 | 141 ± 15 | 0.70 ± 0.13 | 19.2 ± 1.3 | 5.4 ± 1.2 |
| Ni-Al-Cr-W | 4 | 2585 | 18.4 ± 1.1 | 28.2 ± 1.6 | 101 ± 12 | 35 ± 4 | 43 ± 4 |
| | 16 | 1691 | 28.0 ± 1.4 | 43.3 ± 2.2 | 26.7 ± 3.0 | 33 ± 4 | 30 ± 3 |
| | 64 | 1580 | 40.0 ± 1.9 | 57.4 ± 2.8 | 9.9 ± 1.0 | 33 ± 3 | 20.3 ± 1.9 |
| | 264 | 1989 | 56.2 ± 2.8 | 96 ± 9 | 2.7 ± 0.4 | 29.4 ± 1.9 | 13.2 ± 2.5 |

[a.] The mean circular-equivalent precipitate radius ($\langle R \rangle_{PS}$), the mean precipitate three-dimensional radius ($\langle R \rangle$) determined from $PSD_{PS}$ by conversion [44], the number density ($N_v$), volume fraction ($\phi$), and the mean interfacial area per unit volume ($S_v$) with standard errors determined by error propagation [41].

[b.] $N_{ppt}$ is the number of γ'-precipitates analyzed for PSD, which excludes γ'-precipitates intersected by the image edge.

**Figure Captions**

Figure 1. Optical microscopy images of the grain structure for: (a) Ni-9.8 Al-8.3 Cr at.% ; and (b) Ni-9.7 Al-8.5 Cr-2.0 W at.% as revealed by polarized light optical microscopy.

Figure 2. Vickers microhardness versus aging time at 1073 K for the Ni-9.8 Al-8.3 Cr (at.%) and Ni-9.7 Al-8.5 Cr-2.0 W (at.%) alloys.

Figure 4. A typical AFM image taken from a Ni-9.8 Al-8.3 Cr (at.%) sample aged for 64 h at 1073 K

Figure 3. SEM micrographs of the γ'-precipitates (light contrast) protruding from the chemically etched γ-matrix in Ni-9.8 Al-8.3 Cr (at.%) after aging for: (a) 4 h; (b) 16 h; (c) 64 h; and (d) 264 h and in Ni-9.7 Al-8.5 Cr-2.0 W (at.%) after aging for (e) 4 h;  (f) 16 h; (g) 64 h; and (h) 264 h.

Figure 5. SEM micrographs of the γ'-precipitates in Ni-9.8 Al-8.3 Cr (at.%) after aging for (a) 64 h and (b) 264 h, where the polished surface is nearly parallel to an {001}-type plane.

Figure 6. SEM micrographs of the γ'-precipitates in Ni-9.7 Al-8.5 Cr-2.0 W (at.%) after aging for (a) 64 h and (b) 264 h, where the polished surface is nearly parallel to an {001}-type plane.

Figure 7. The stereologically determined volume fraction measurements compared to the phase extraction (PE) measurements, denoted by the solid line, for the 264 h aging state of the Ni-9.8 Al-8.3 Cr (at.%) and Ni-9.7 Al-8.5 Cr-2.0 W (at.%) alloys aged isothermally at 1073 K.

Figure 8. The temporal evolution of the mean precipitate radius and interfacial area per unit volume for a Ni-9.8 Al-8.3 Cr (at.%) alloy aged at 1073 K. The coarsening rate constants are determined from the best linear fit of the data.

Figure 9. The temporal evolution of the mean γ'-precipitate radius and interfacial area per unit volume for a Ni-9.7 Al-8.5 Cr-2.0 W (at.%) alloy aged at 1073 K. The coarsening rate constants are determined from the best linear fit of the data.

Figures 10. The temporal evolution of the number density of γ'-precipitates in Ni-9.8 Al-8.3 Cr and Ni-9.7 Al-8.5 Cr-2.0 W (at.%) alloys aged isothermally at 1073 K.

Figure 11. Temporal evolution of the scaled γ'-precipitate size distributions for a Ni-9.8 Al-8.3 Cr at.% alloy aged at 1073 K with $\phi^{eq} = 0.189$, compared to the stationary-state prediction of Akaiwa and Voorhees [27] for $\phi = 0.2$.

Figure 12. Temporal evolution of the scaled γ'-precipitate size distributions for a Ni-9.7 Al-8.5 Cr-2.0 W at.% alloy aged at 1073 K with $\phi^{eq} = 0.308$ compared to the stationary-state prediction of Akaiwa and Voorhees [27] for $\phi = 0.3$.

Figure 13. Temporal evolution of the normalized pair correlation function for (a) Ni-9.8 Al-8.3 Cr (at.%) and (b) Ni-9.7 Al-8.5 Cr-2.0 W (at.%) alloys aged at 1073 K.

Figure 14. Temporal evolution of edge-to-edge interprecipitate distances for Ni-9.8 Al-8.3 Cr (at.%) alloy aged at 1073 K, where both (a) frequency distribution and (b) normalized distribution scaled to mean circular-equivalent radius are plotted.

Figure 15. Temporal evolution of edge-to-edge interprecipitate distances for Ni-9.7 Al-8.5 Cr-2.0 W (at.%) alloy aged at 1073 K, where both (a) frequency distribution and (b) normalized distribution scaled to mean circular-equivalent radius are plotted.

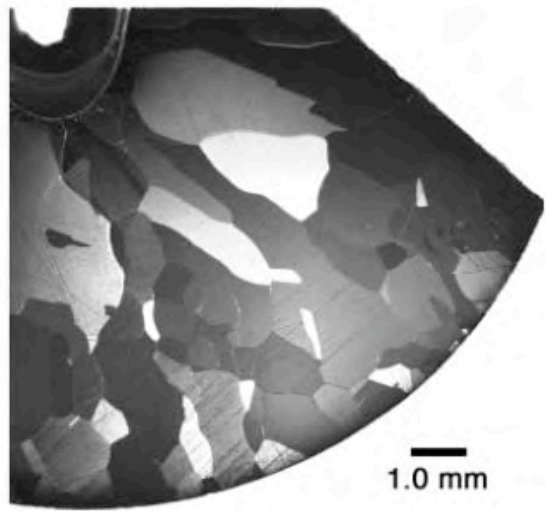

**(a)**

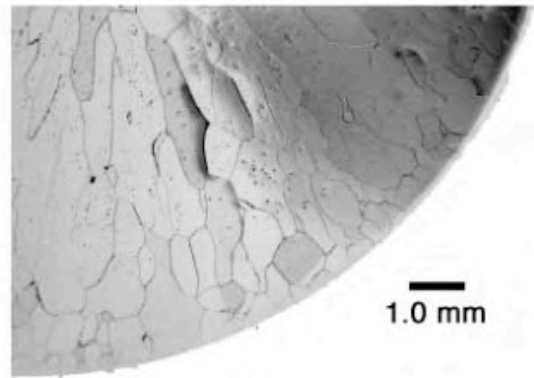

**(b)**

Figure 1

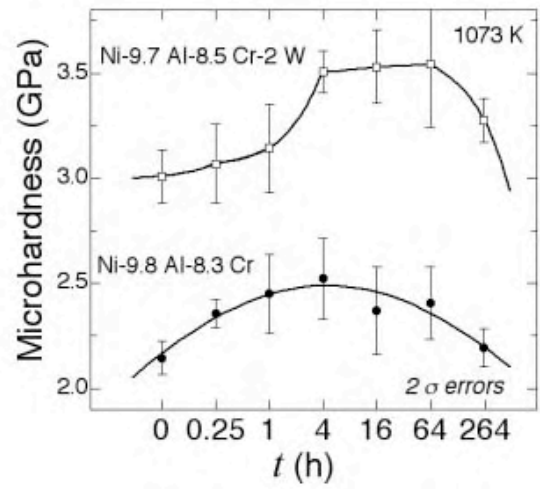

Figure 2

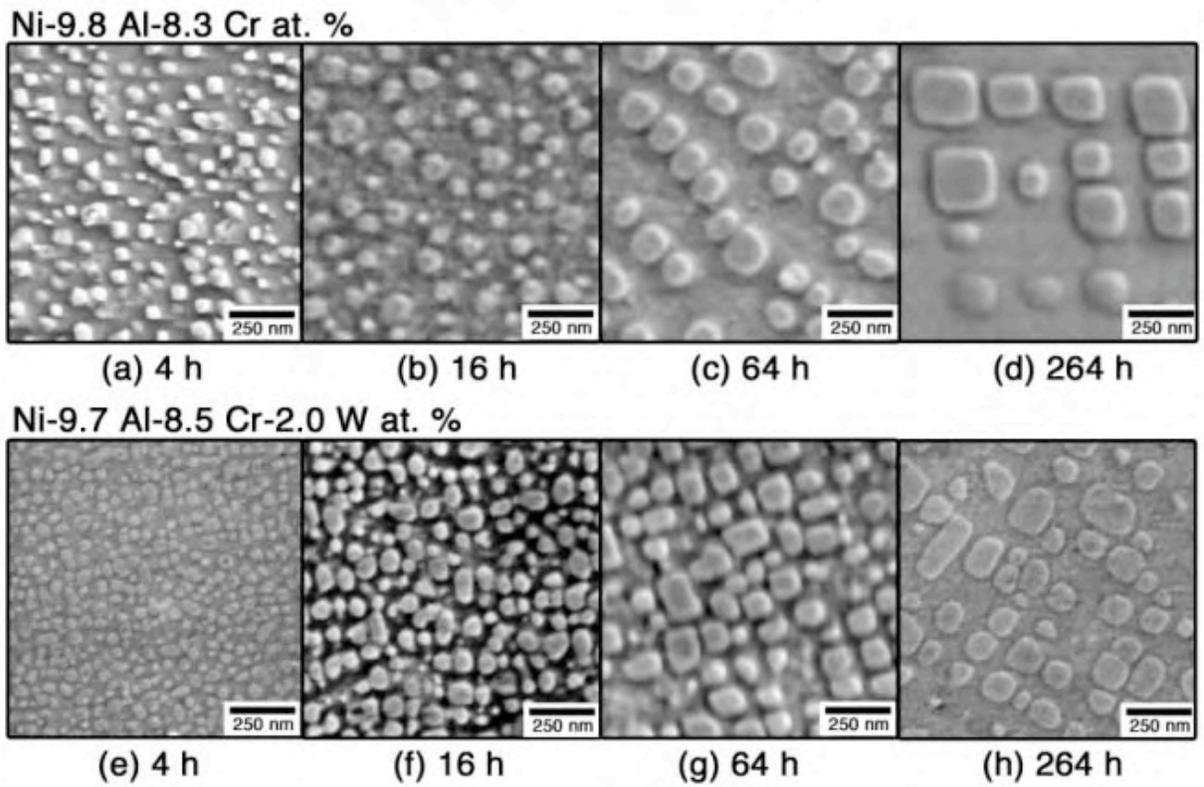

Figure 3

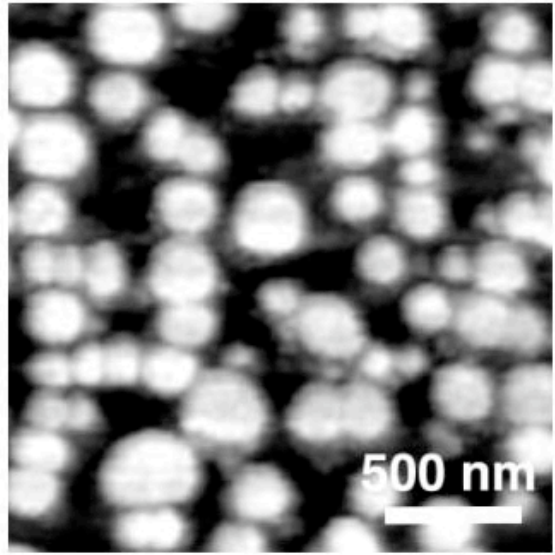

Figure 4

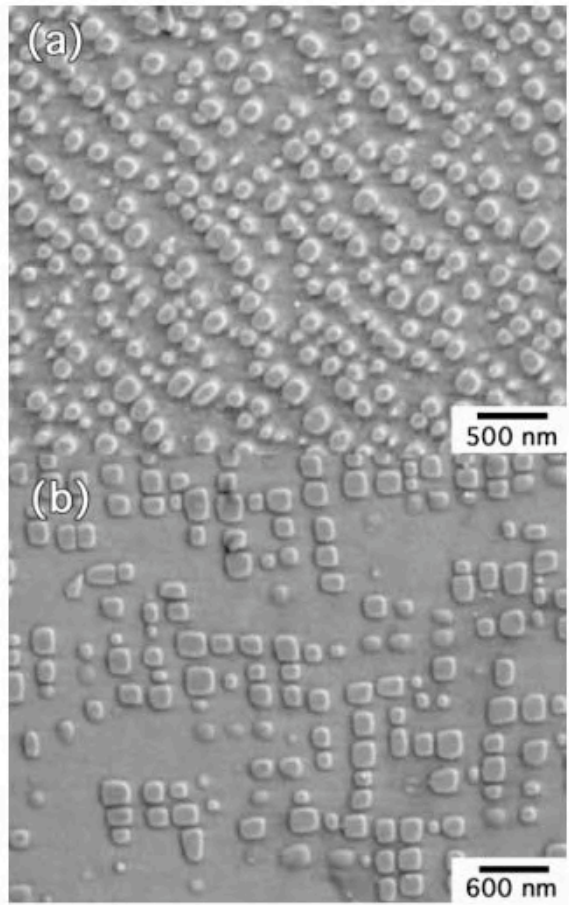

Figure 5

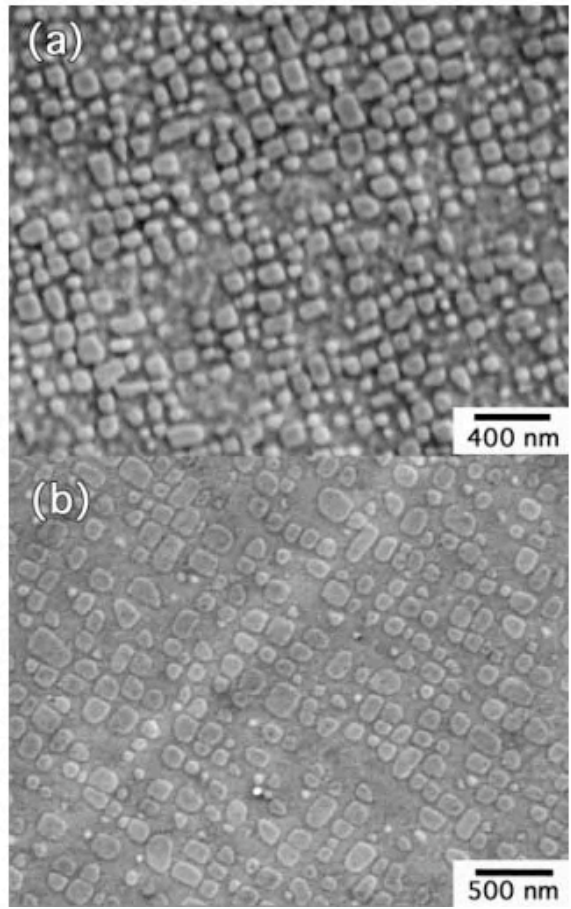

Figure 6

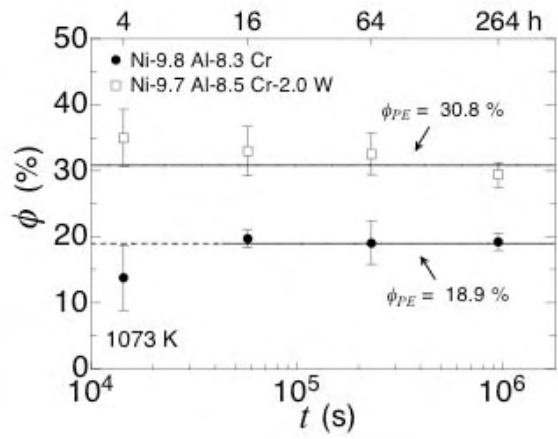

Figure 7

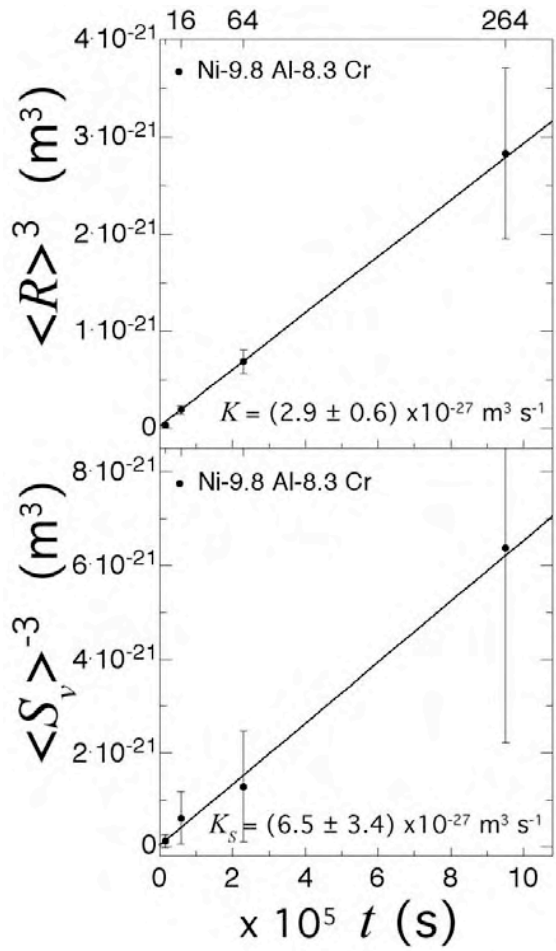

Figure 8

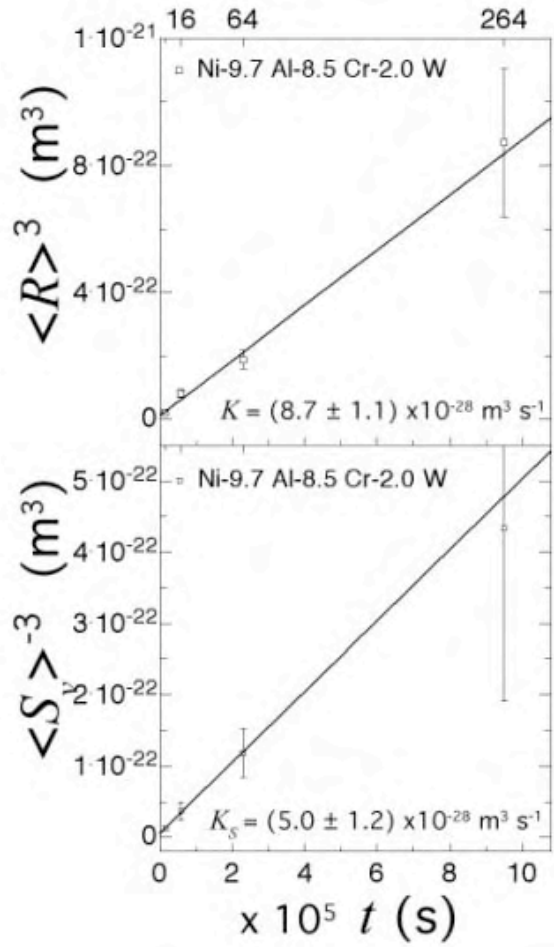

Figure 9

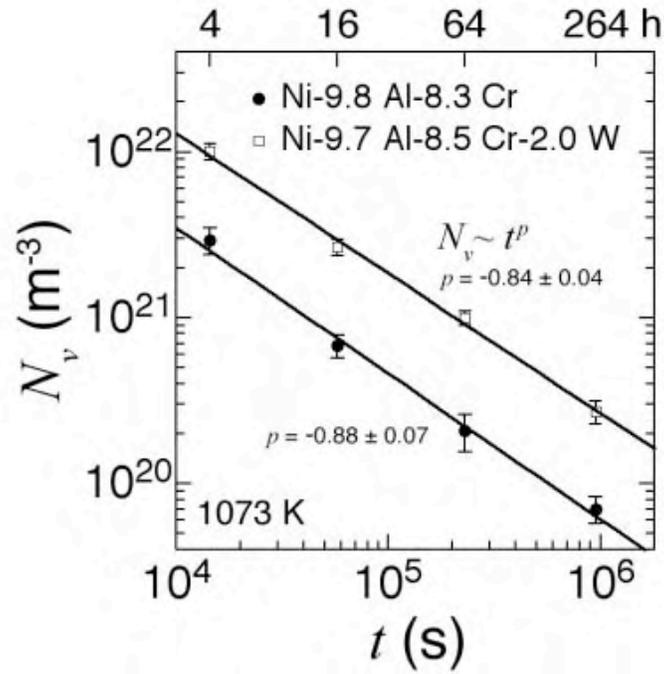

Figure 10

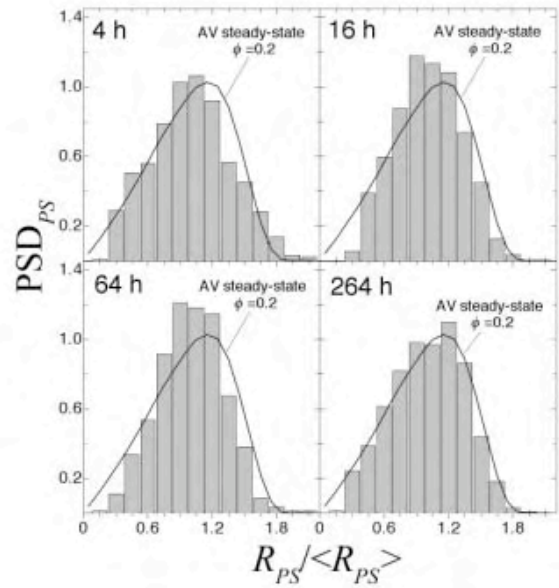

Figure 11

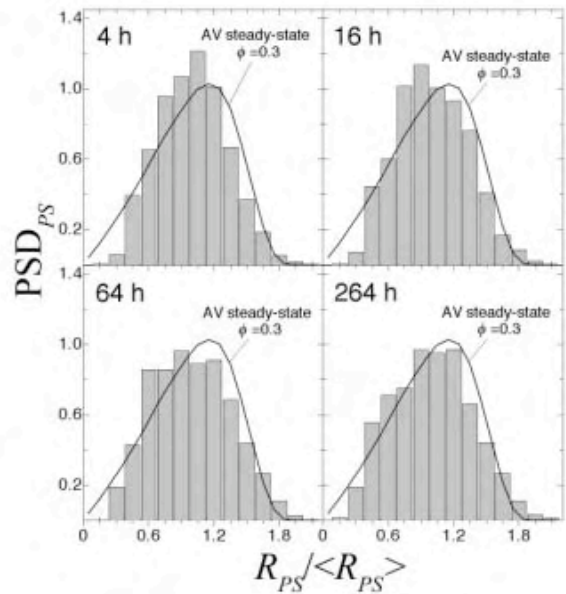

Figure 12

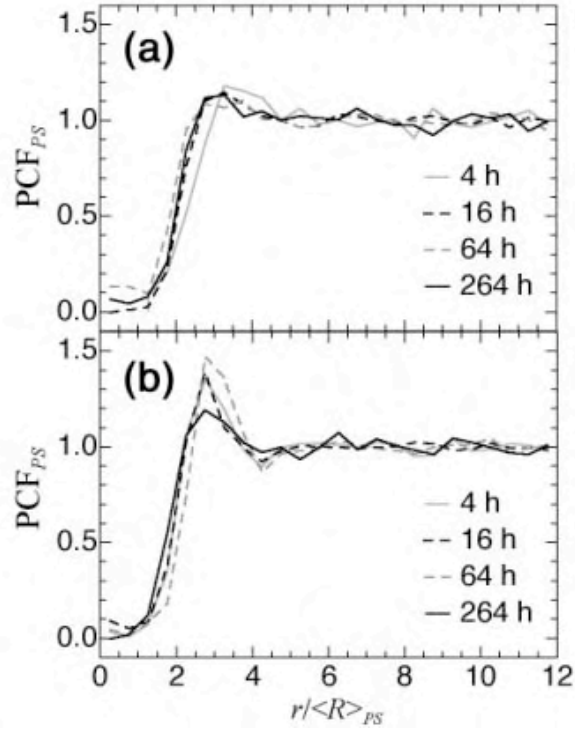

Figure 13

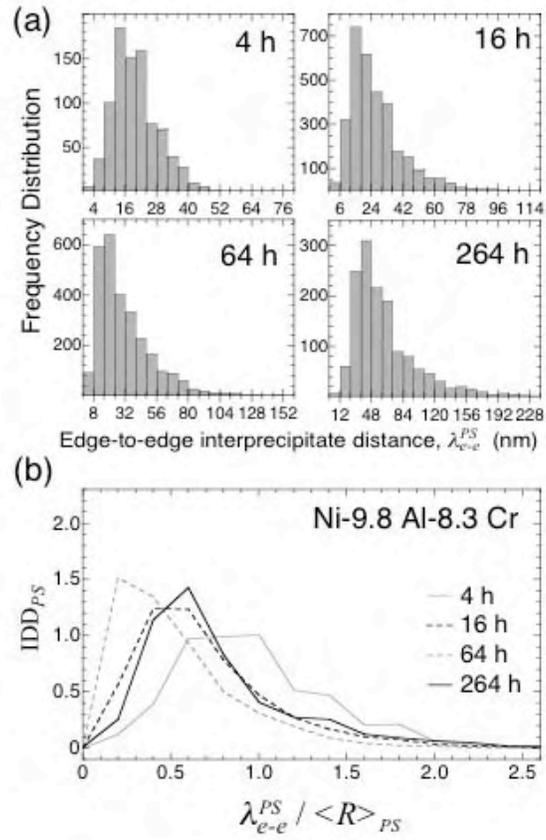

Figure 14

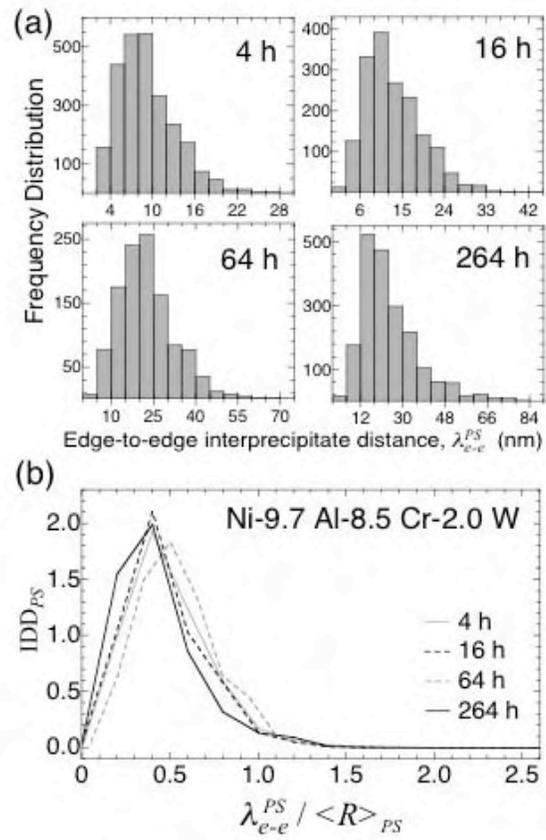

Figure 15